\documentclass{ieeetran}
\usepackage{amssymb}
\usepackage{amsmath}
\usepackage{graphicx}
\usepackage{slashbox}
\usepackage{array}
\usepackage{color}
\usepackage{amsfonts}

\newtheorem{Theorem 1}{Theorem}
\newtheorem{Theorem 2}[Theorem 1]{Theorem}
\newtheorem{Theorem 3}[Theorem 1]{Theorem}
\newtheorem{Theorem 4}[Theorem 1]{Theorem}
\newtheorem{Theorem 5}[Theorem 1]{Theorem}
\newtheorem{Theorem 6}[Theorem 1]{Theorem}
\newtheorem{Theorem 7}[Theorem 1]{Theorem}
\newtheorem{Theorem 8}[Theorem 1]{Theorem}
\newtheorem{Assumption 1}{Assumption}
\newtheorem{Assumption 2}[Assumption 1]{Assumption}
\newtheorem{Assumption 3}[Assumption 1]{Assumption}
\newtheorem{Assumption 4}[Assumption 1]{Assumption}
\newtheorem{Assumption 5}[Assumption 1]{Assumption}
\newtheorem{Definition 1}{Definition}
\newtheorem{Remark 1}{Remark}
\newtheorem{Remark 2}[Remark 1]{Remark}
\newtheorem{Remark 3}[Remark 1]{Remark}
\newtheorem{Remark 4}[Remark 1]{Remark}
\newtheorem{Remark 5}[Remark 1]{Remark}
\newtheorem{Remark 6}[Remark 1]{Remark}
\newtheorem{Remark 7}[Remark 1]{Remark}
\newtheorem{Remark 8}[Remark 1]{Remark}
\newtheorem{Remark 9}[Remark 1]{Remark}
\newtheorem{Remark 10}[Remark 1]{Remark}
\newtheorem{Remark 11}[Remark 1]{Remark}
\newtheorem{Lemma 1}{Lemma}
\newtheorem{Lemma 2}[Lemma 1]{Lemma}
\newtheorem{Lemma 3}[Lemma 1]{Lemma}

\begin{document}
\title{Exponential synchronization rate of Kuramoto oscillators in the presence of a pacemaker}
\author{Yongqiang Wang, {\it Member, IEEE}, Francis
J. Doyle III, {\it Fellow, IEEE}\thanks{The work was supported in
part by U.S. ARO (W911NF-07-1-0279), NIH (GM078993), and ICB
(W911NF-09-0001) from the U.S. ARO. The content of the information
does not necessarily reflect the position or the policy of the
Government, and no official endorsement should be inferred.}
\thanks{Yongqiang Wang, Francis J. Doyle III are with the Institute for Collaborative Biotechnologies, University of California, Santa
Barbara,  California 93106-5080 USA. E-mail: wyqthu@gmail.com,
frank.doyle@icb.ucsb.edu.}} \maketitle
\begin{abstract}
 The exponential synchronization rate is addressed
for Kuramoto oscillators in the presence of a pacemaker.  When
natural frequencies are identical,  we prove that synchronization
can be ensured even when the phases are not constrained in an open
half-circle, which improves the existing results in the literature.
We derive a lower bound on the exponential synchronization rate,
which is proven to be an increasing function of pacemaker strength,
but may be an increasing or decreasing function of local coupling
strength. A similar conclusion is obtained for phase locking when
the natural frequencies are non-identical. An approach to trapping
phase differences in an arbitrary interval is also given, which
ensures  synchronization
 in the sense that synchronization error can be reduced
to an arbitrary level.
\end{abstract}

\section{Introduction}

The Kuramoto model was first proposed in 1975 to model the
synchronization of chemical oscillators sinusoidally coupled in an
all-to-all architecture \cite{kuramoto:75}. Although it is elegantly
simple, the Kuramoto model is sufficiently flexible to be adapted to
many different contexts, hence it is widely used and is regarded as
one the most representative models of coupled phase oscillators
\cite{acebron:05}. Recently, the Kuramoto model has
received increased attention. 
For example, the authors in \cite{chopra:09,Scardovi:07,
Verwoerd:07} discussed synchronization conditions for the Kuramoto
model. The work in \cite{Papachristodoulou:05} gave a
synchronization condition for delayed  Kuramoto oscillators. Results
are also obtained for Kuramoto oscillators with coupling topologies
different from the original all-to-all structure. For example, the
authors in \cite{Rogge:04} and  \cite{Klein:09} considered the phase
locking of Kuramoto oscillators coupled in a ring and a chain,
respectively. Using graph theory, the authors in \cite{Jadbabaie:04,
Lin:07,Papachristodoulou:10} discussed the synchronization of
Kuramoto oscillators with arbitrary coupling topologies.
  The authors in
\cite{Dorfler:11} proved that exponential
 synchronization can be achieved for Kuramoto oscillators when phases lie in an open half-circle.

Studying the influence of the pacemaker (also called the leader, or
the pinner \cite{Delellis:11}) on Kuramoto oscillators is not only
of theoretical interest, but also of practical importance
\cite{Childs:08, wang_automatica:11}. For example, in circadian
systems,  thousands of clock cells in the brain are entrained to the
light-dark cycle \cite{Herzog:07}. In the
 clock synchronization
of wireless networks,  time references in individual nodes are
synchronized by means of intercellular interplay and external
coordination from a time base such as GPS \cite{Kopetz:87}. Hence,
Kuramoto oscillators  with a pacemaker are attracting increased
attention. The authors in \cite{Childs:08} and \cite{sakaguchi:88}
studied the bifurcation diagram and the steady macroscopic rotation
of Kuramoto oscillators forced by a pacemaker that acts on every
node. Based on numerical methods, the authors in \cite{Kori:04}
showed that the network depth (defined as the mean distance of nodes
from the pacemaker, a term closely related to
pinning-controllability in pinning control \cite{Porfiri:08})
affects the entrainment of randomly coupled Kuramoto oscillators to
a pacemaker. Using numerical methods, the authors in
\cite{popovych:11} discovered that there may be situations in which
the population field potential is entrained to the pacemaker while
individual oscillators are phase desynchronized. But compared with
the rich results on pacemaker-free Kuramoto oscillators, analytical
results are relative sparse for Kuromoto oscillators forced by a
 pacemaker. And to our knowledge, there are no existing
results on the synchronization rate of arbitrarily coupled Kuramoto
oscillators  in the presence of a pacemaker.

The synchronization rate is crucial in many synchronized processes.
For example, in the main olfactory system, stimulus-specific
ensembles of neurons synchronize their firing to  facilitate odor
discrimination, and the synchronization time determines the speed of
olfactory discrimination \cite{Uchida:03}. In the clock
synchronization of wireless sensor networks, the synchronization
rate is a determinant of energy consumption, which is vital for
cheap sensors \cite{wang_tsp:12,wang_tcst:12}.

We consider the exponential synchronization rate of
 Kuramoto oscillators with an arbitrary topology in the
presence of a pacemaker. In the identical natural frequency case, we
prove that synchronization (oscillations with identical phases) can
be ensured, even when phases are not constrained in an open
half-circle. In the non-identical natural frequency case where
perfect synchronization has been shown cannot be achieved
\cite{acebron:05},\cite{Strogatz:00}, we prove that phase locking
(oscillations with  identical oscillating frequencies) can be
ensured and synchronization can be achieved in the sense that phase
differences can be reduced to an arbitrary level. In both cases, the
influences of the pacemaker and local coupling strength on the  synchronization rate are analyzed. 
%

\section{Problem formulation and Model transformation}\label{sec:problem
formulation}
Consider a network of  $N$ oscillators, which will henceforth be
referred to as 'nodes'. All $N$ nodes (or a subset) receive
alignment information from a pacemaker (also called the leader, or
the pinner \cite{Delellis:11}). Denoting the phases of the pacemaker
and  node $i$ as $\varphi_0$ and  $\varphi_i$, respectively,
 the  dynamics of the Kuramoto oscillator network can be
written as
\begin{eqnarray}\label{eq:model_varphi}
\left\{
\begin{aligned}
\dot\varphi_0&=w_0 \\
\dot\varphi_i&=w_i+ \hspace{-0.2cm}\sum\limits_{j=1,j\neq
i}^{N}\hspace{-0.2cm}a_{i,j}\sin\left(\varphi_j-\varphi_i\right)+g_i\sin\left(\varphi_0-\varphi_i\right)
\end{aligned}
\right.
\end{eqnarray}
for $1\leq i\leq N $,
 where $w_0$ and $w_i$ are the natural frequencies of the pacemaker
 and the $i$th oscillator, respectively,
$a_{i,j}\sin\left(\varphi_j-\varphi_i\right)$ is the interplay
between node $i$ and node $j$ with  $a_{i,j}\geq 0$ denoting the
strength, $g_i\sin(\varphi_0-\varphi_i)$ denotes the force of the
pacemaker with $g_i\geq0$ denoting its strength. If $a_{i,j} = 0$
(or $g_i=0$), then oscillator $i$ is not influenced by oscillator
$j$ (or the pacemaker).

\begin{Assumption 1}\label{assump:assumption 1}
We assume symmetric coupling between pairs of oscillators, i.e.,
$a_{i,j}=a_{j,i}$.
\end{Assumption 1}

%

Next, we study the influences of the pacemaker, $g_{i}$, and local
coupling, $a_{i,j}$, on the
rate of exponential synchronization. 

Solving the first equation in (\ref{eq:model_varphi}) gives the
dynamics of the pacemaker $\varphi_0=w_0t+\phi_0$, where the
constant $\phi_0$ denotes the initial phase. To study if oscillator
$i$ is synchronized to the pacemaker, it is convenient to study the
phase deviation of oscillator $i$ from the pacemaker. So we
introduce the following change of variables:
\begin{eqnarray}\label{eq:xi}
\varphi_i=\varphi_0+\xi_i=w_0t+\phi_0+\xi_i
\end{eqnarray}
$\xi_i\in[-2\pi,\,2\pi]$ denotes the phase deviation of the $i$th
oscillator from the pacemaker. Due to the $2\pi$-periodicity of the
sine-function, we can restrict our attention to
$\xi_i\in[-\pi,\,\pi]$.
 Substituting (\ref{eq:xi}) into
(\ref{eq:model_varphi}) yields the dynamics of $\xi_i$:
\begin{eqnarray}\label{eq:model_xi}
\begin{aligned}
\dot\xi_i=w_i-w_0+\sum\limits_{j=1,j\neq
i}^Na_{i,j}\sin\left(\xi_j-\xi_i\right)-g_i\sin(\xi_i)
\end{aligned}
\end{eqnarray}

Since $\xi_i$ is the relative phase of the $i$th oscillator with
respect to the phase of the pacemaker, it will be referred to as
relative phase in the remainder of the paper.

By studying the properties of (\ref{eq:model_xi}), we can obtain:
\begin{itemize}
\item Condition for synchronization: If all $\xi_i$
 converge to $0$, then we have
 $\varphi_1=\varphi_2=\ldots=\varphi_N=\varphi_0$ when $t\rightarrow\infty$, meaning that all
nodes are synchronized to the pacemaker.
\item Exponential synchronization rate: The rate of synchronization is
determined by the rate at which $\xi_i$ decays to $0$, namely, it
 can be measured by the maximal $\alpha$  satisfying
\begin{equation}
\|\xi(t)\|\leq Ce^{-\alpha t}\|\xi(0)\|,\quad \xi=[
\xi_1,\,\xi_2,\,\ldots,\,\xi_N]^T
\end{equation}
for some constant $C$, where $\|\bullet\|$ is the Euclidean norm.
$\alpha$ measures the exponential synchronization rate of
(\ref{eq:model_xi}): a larger $\alpha$ leads to a faster
synchronization rate.
\end{itemize}
\begin{Remark 1}
When  $w_i$ and $w_0$  are non-identical, synchronization
($\xi_i=0$) cannot be achieved in general. But we will prove in Sec.
\ref{se:phase trapping} that the synchronization error  can be made
arbitrarily small by tuning the strength of the pacemaker $g_i$.
\end{Remark 1}

Assigning arbitrary orientation to each interaction, we can get the
$N\times M$ incidence matrix  $B$  ($M$ is the number of interaction
edges, i.e., non-zero $a_{i,j}\,(1\leq i\leq N,\,j<i)$) of the
interaction graph \cite{Godsil:01}: $B_{i,j}=1$ if edge $j$ enters
node $i$, $B_{i,j}=-1$ if edge $j$ leaves node $i$, and $B_{i,j}=0$
otherwise. Then using graph theory, (\ref{eq:model_xi}) can be
recast in a matrix form:
\begin{equation}\label{eq:matrix form}
\dot\xi=\Omega -G\sin\xi-BW\sin\left(B^T\xi\right)
\end{equation}
where $\Omega=[w_1-w_0,\,w_2-w_0,\,\ldots,\,w_N-w_0]^T$, $
G=\textrm{diag}(g_1,\:g_2,\:\ldots,\:g_N)$, and
$W=\textrm{diag}(\nu_1,\:\nu_2,\:\ldots,\: \nu_M)$. Here
$\nu_i\,(1\leq i\leq M)$ are  a permutation of non-zero
$a_{i,j}\,(1\leq i\leq N,j<i)$ and $\textrm{diag}(\bullet)$ denotes
a diagonal matrix.

\section{The identical natural frequency case}\label{se:synch rate under identical
frequencies} When $w_1=w_2=\ldots=w_N=w_0$, (\ref{eq:matrix form})
reduces to:
\begin{equation}\label{eq:matrix form_identical case}
\dot\xi=-G\sin\xi-BW\sin\left(B^T\xi\right)
\end{equation}

To study the exponential synchronization rate, we first give a
synchronization condition:
\begin{Theorem 1}\label{theo:Theorem 1}
For the network  in (\ref{eq:matrix form_identical case}), denote
$\varepsilon\triangleq \max\limits_{1\leq i\leq N}|\xi_i|$ and ${\rm
sinc}(x) \triangleq\sin(x)/x$, then
\begin{enumerate}
\item when $\varepsilon<\frac{\pi}{2} $, the
network synchronizes if at least one $g_i$ is positive and the
coupling $a_{i,j}$ is connected, i.e., there is a multi-hop link
from each node to every other node;
\item when $\frac{\pi}{2}\leq\varepsilon<\pi$, the network
  synchronizes if the following inequality is satisfied:
\begin{equation}\label{eq:condition_2}
\begin{aligned}
&\hspace{-0.15cm}g_{\min}>&\\
&\hspace{-0.15cm}\max\hspace{-0.1cm}\left\{\hspace{-0.10cm}\frac{{\rm
sinc} (2\varepsilon_0)\lambda_{\max}(BWB^T)}{-{\rm
sinc}(\varepsilon)},\:
\max\limits_i\big\{\hspace{-0.20cm}\sum\limits_{j=1,j\neq i}^N\frac{
a_{i,j}}{\sin(\varepsilon)}\big\}\right\}&
\end{aligned}
\end{equation}
where  $\lambda_{\max}(\bullet)$ denotes the maximal eigenvalue,
$g_{\min}$ and $\varepsilon_0\in(\frac{\pi}{2},\,\pi)$ are
determined by
\begin{eqnarray}\label{eq:tilde}
\hspace{-0.2cm}g_{\min}=\min\{g_1,\ldots,g_N\}, \,2\varepsilon_0
\cos\left(2\varepsilon_0\right)=\sin(2\varepsilon_0)
\end{eqnarray}
\end{enumerate}
\end{Theorem 1}
\begin{proof}
We first prove that when  $\xi\in [-\varepsilon,\,\varepsilon]\times
\ldots\times[-\varepsilon,\,\varepsilon]=[-\varepsilon,\,\varepsilon]^N$
where $\times$ denotes Cartesian product, they will remain in the
interval under conditions in Theorem \ref{theo:Theorem 1}, i.e, the
$n$-tuple set $[-\varepsilon,\,\varepsilon]^N$ is positively
invariant for (\ref{eq:matrix form_identical case}).

To prove the positive invariance of
$[-\varepsilon,\,\varepsilon]^N$, we only need to check the
direction of vector field on the boundaries. When $\varepsilon<
\frac{\pi}{2}$, if $\xi_i=\varepsilon$, we have
$-\pi<-2\varepsilon\leq\xi_j-\xi_i\leq0$ for $1\leq j\leq N$. So in
(\ref{eq:model_xi}), $\sin(\xi_j-\xi_i)\leq0$ and $\sin(\xi_i)>0$
hold, and hence $\dot{\xi}_i<0$ holds (Note that  $w_i-w_0=0$).
Hence the vector field is pointing inward in the set, and no
trajectory can escape to values larger than $\varepsilon$.
Similarly, we can prove that when $\xi_i=-\varepsilon$,
$\dot{\xi}_i> 0$ holds. Thus no trajectory can escape to values
smaller than $-\varepsilon$. Therefore $
[-\varepsilon,\,\varepsilon]^N$ is positively invariant when
$\varepsilon<\frac{\pi}{2}$. When $\frac{\pi}{2}\leq
\varepsilon<\pi$, if $\xi_i=\varepsilon$, we have
$\sin\xi_i=\sin\varepsilon>0$ and $\sin(\xi_j-\xi_i)\leq 1$ for
$1\leq j\leq N$. So when $w_i=w_0$, if (\ref{eq:condition_2}) is
satisfied, the right hand side of (\ref{eq:model_xi}) is negative,
i.e., $\dot{\xi}_i< 0$ holds. Therefore the vector field is pointing
inward in the set and no trajectory can escape to values larger than
$\varepsilon$. Similarly, we can prove that if $\xi_i=-\varepsilon$,
$\dot{\xi}_i> 0$ holds under condition (\ref{eq:condition_2}). Thus
no trajectory can escape to values smaller than $-\varepsilon$.
Therefore $ [-\varepsilon,\,\varepsilon]^N$ is also positively
invariant for $\frac{\pi}{2}\leq \varepsilon<\pi$ if
(\ref{eq:condition_2}) is satisfied.

Next we proceed to prove synchronization.  Define a Lyapunov
function as $ V=\frac{1}{2}\xi^T\xi $. $V\geq 0$ is zero iff all
$\xi_i$ are zero, meaning the synchronization of all nodes to the
pacemaker.

Differentiating $V$ along the trajectories of (\ref{eq:matrix
form_identical case}) yields
\begin{equation}\label{eq:dot V}
\begin{aligned}
\dot{V}=\xi^T\dot{\xi}&=-\xi^T\left(G\sin\xi+BW\sin(B^T\xi)\right)\\
&=-\xi^TGS_1\xi-\xi^TBWS_2B^T\xi
\end{aligned}
\end{equation}
where $S_1\in\mathcal{R}^{N\times N}$ and
$S_2\in\mathcal{R}^{M\times M}$ are given by
\begin{equation}\label{eq:S1 and S2}
\begin{aligned}
S_1&=\textrm{diag}\left\{{\rm sinc}(\xi_1),\,\ldots,\,{\rm
sinc}(\xi_N)\right\},\\
 S_2&=\textrm{diag}\left\{ {\rm sinc}
(B^T\xi)_1,\,\dots,\,{\rm sinc} (B^T\xi)_M\right\}
\end{aligned}
\end{equation}
%
%
with  $(B^T\xi)_i$ denoting the $i$th element of the $M\times 1$
dimensional vector $B^T\xi$.

From dynamic systems theory, if $GS_1+BWS_2B^T$ in (\ref{eq:dot V})
is positive definite when $\xi\neq 0$, then $\dot{V}$ is negative
when $\xi\neq0$ and $V$ will decay to zero, meaning that $\xi$ will
decay to zero and all nodes are synchronized to the pacemaker.

1) When all $\xi_i$ are within $[-\varepsilon,\,\varepsilon]$ with
$0\leq\varepsilon<\frac{\pi}{2}$, $(B^T\xi)_i$ is in the form of
$\xi_m-\xi_n\,(1\leq m,n\leq N)$, and hence is restricted to
$(-\pi,\,\pi)$.  Given that in $(-\pi,\,\pi)$,  $ {\rm sinc}(x)> 0 $
holds, it follows that $S_1$ and $S_2$ satisfy the following
inequalities:
\begin{eqnarray}\label{eq:sigma_1 and sigma_2}
\begin{aligned}
S_1&\geq \sigma_1 I, \quad
\sigma_1\triangleq\min\limits_{-\varepsilon\leq
x\leq\varepsilon}{\rm sinc}(x)={\rm sinc}(\varepsilon),\\
S_2&\geq \sigma_2 I,\quad
\sigma_2\triangleq\min\limits_{-2\varepsilon\leq
x\leq2\varepsilon}{\rm sinc}(x)={\rm sinc}(2\varepsilon)
\end{aligned}
\end{eqnarray}
So we have $GS_1+BWS_2B^T\geq \sigma_1G+\sigma_2BWB^T$, which in
combination with (\ref{eq:dot V}) produces
\begin{equation}
\dot{V}\leq -\xi^T\left(\sigma_1G+\sigma_2BWB^T \right)\xi
\end{equation}

It can be verified that $\sigma_1G+\sigma_2BWB^T$ is of form:
\begin{equation}\label{eq:explicit formulation of Lambda}
\sigma_1G+\sigma_2BWB^T=\sigma_1{\rm
diag}\{g_1,\,g_2,\,\ldots,\,g_N\}+\sigma_2L
\end{equation}
with$\,L\in\mathcal{R}^{N\times N}$given as follows: for $i\neq j$,
its $(i,j)$th element is $-a_{i,j}$, for $i=j$, its $(i,j)$th
element is $\sum_{ m=1, m\neq i}^N a_{i,m}$.
Since $\sigma_1$ and $\sigma_2$ are positive,  $g_i$ and $a_{i,j}$
are non-negative, it follows from the Gershgorin Circle Theorem that
$\sigma_1G+\sigma_2BWB^T$ only has  non-negative eigenvalues
\cite{horn:85}. Next we prove its positive definiteness by excluding
 $0$ as an eigenvalue.

Since the topology of  $a_{i,j}$ is connected,
$\sigma_1G+\sigma_2BWB^T$ is irreducible from graph theory
\cite{horn:85}. This in combination with the assumption of at least
one $g_i>0$ guarantees that $\sigma_1G+\sigma_2BWB^T$ is irreducibly
diagonally dominant. So from Corollary 6.2.27 of \cite{horn:85}, we
know its determinant is non-zero, and hence 0 is not its eigenvalue.
Therefore $\sigma_1G+\sigma_2BWB^T$ is positive definite, and hence
$V$ will converge to $0$, meaning that the nodes will synchronize to
the pacemaker.

2) When $\xi_i\in[-\varepsilon,\,\varepsilon]$ ($1\leq i\leq N$)
with $\frac{\pi}{2}\leq\varepsilon<\pi$, $S_1$ is positive definite
but $S_2$ is not  since $(B^T\xi)_i$ is in
$[-2\varepsilon,\,2\varepsilon]$,  and thus ${\rm sinc}(B^T\xi)_i$
may be negative. It can be proven that ${\rm sinc }(x)$ is
monotonically decreasing on $[0,\,2\varepsilon_0]$ and monotonically
increasing on $[2\varepsilon_0,\,2\pi]$ (using the first derivative
test), where $\varepsilon_0\in(\frac{\pi}{2},\,\pi)$ is determined
by (\ref{eq:tilde}). Hence we have $ S_1\geq {\rm
sinc}(\varepsilon)I$ and $S_2\geq {\rm sinc}(2\varepsilon_0)I$ where
${\rm sinc}(2\varepsilon_0)<0$. Therefore (\ref{eq:dot V}) reduces
to
\begin{equation}\label{eq:V in theorem 2}
\begin{aligned}
\dot{V}&\leq-{\rm sinc}(\varepsilon)\xi^TG\xi-{\rm
sinc}(2\varepsilon_0)\xi^TBWB^T\xi \\
&\leq -\xi^T\left({\rm sinc}(\varepsilon) G+{\rm
sinc}(2\varepsilon_0)BWB^T\right)\xi
\end{aligned}
\end{equation}
Thus $\xi\rightarrow 0$ if $ g_{\min}{\rm sinc}(\varepsilon)+{\rm
sinc}(2\varepsilon_0)\lambda_{\max}(BWB^T)>0$ holds.
\end{proof}

\begin{Remark 1}
It is already known that for general Kuramoto oscillators without a
pacemaker, synchronization can only be ensured when
$\max\limits_i{\varphi_i}-\min\limits_i{\varphi_i}$ is less than
$\pi$, i.e., the initial phases lie in an open half-circle
\cite{Jadbabaie:04,Lin:07,Papachristodoulou:10,Dorfler:11,Monzon:05,wang_tsp2:12}
(although when phases are lying outside a half-circle, {\it almost
global synchronization} is possible by replacing the sinusoidal
interaction function with elaborately designed periodic functions
\cite{Mallada:10,Sarlette:09}, it may introduce numerous unstable
equilibria \cite{Sarlette:09}).  Here, synchronization is ensured
even when $\xi_i$ is outside $(-\frac{\pi}{2},\,\frac{\pi}{2})$,
i.e., when phase difference $\varphi_i-\varphi_j=\xi_i-\xi_j$ is
larger than $\pi$, meaning that the phases can lie outside a
half-circle. This shows the advantages of introducing a pacemaker.
\end{Remark 1}
\begin{Remark 2}
Theorem \ref{theo:Theorem 1} indicates that when $\xi_i$ is outside
$(-\frac{\pi}{2},\,\frac{\pi}{2})$, i.e., when phases cannot be
constrained in one open half-circle, all nodes have to be connected
to the pacemaker to ensure synchronization. In fact, when some
oscillators are not connected to the pacemaker, the relative phases
may not converge to 0. For example, consider two connected
oscillators,  1 and  2, with coupling strength
$a_{1,2}=a_{2,1}=\kappa$. If the pacemaker only acts on oscillator 1
with strength $g_1= \kappa$ and the phases of the
 pacemaker, oscillator 1 and  2 are $\pi$, $0.4\pi$, and $1.6\pi$, respectively, though
 $\xi_1=-0.6\pi$ and $\xi_2=0.6\pi$ are all within
 $[-0.6\pi,\,0.6\pi]$,
 numerical simulation shows that  $\xi_2$ will not converge to 0 no matter how large $\kappa$ is.
\end{Remark 2}
\begin{Remark 6}
Since the eigenvalues of $BWB^T$ are non-negative \cite{Godsil:01},
 $\lambda_{\max}(BWB^T)>0$.
\end{Remark 6}

Based on a similar derivation, we can get a bound on the exponential
synchronization rate:
\begin{Theorem 2}\label{theo:Theorem 2}
For the network  in (\ref{eq:matrix form_identical case}), denote
$\varepsilon\triangleq \max\limits_{1\leq i\leq N}|\xi_i|$. If the
 conditions in Theorem \ref{theo:Theorem 1} are
satisfied, then the exponential synchronization rate can be bounded
as follows:
\begin{enumerate}
\item when
$0\leq\varepsilon<\frac{\pi}{2}$ holds, the exponential
synchronization rate is no worse than
\begin{equation}\label{eq:alpha_1}
\begin{aligned}
\alpha_1&=\min\limits_{\xi}\left\{ \xi^T
\left(\sigma_1G+\sigma_2BWB^T\right)\xi /(\xi^T\xi)\right\}\\
&=\lambda_{\min}\left(\sigma_1G+\sigma_2BWB^T\right)
\end{aligned}
\end{equation}
with $\sigma_1G+\sigma_2BWB^T$ given in (\ref{eq:explicit
formulation of Lambda});

\item when
$\frac{\pi}{2}\leq\varepsilon<\pi$ holds, the exponential
synchronization rate is no worse than
\begin{equation}\label{eq:alpha_2}
\alpha_2=g_{\min}{\rm sinc}(\varepsilon)+{\rm
sinc}(2\varepsilon_0)\lambda_{\max}(BWB^T)
\end{equation}
\end{enumerate}
\end{Theorem 2}
\begin{proof}
From the proof in Theorem \ref{theo:Theorem 1}, when
$0\leq\varepsilon<\frac{\pi}{2}$, we have $ \dot{V}\leq -2\alpha_1 V
$, which means $ V(t)\leq C^2e^{-2\alpha_1 t}V(0)\Rightarrow
\|\xi(t)\|\leq C e^{-\alpha_1 t}\|\xi(0)\| $  for some positive
constant $C$. Thus the  synchronization rate is no less than
$\alpha_1$.

Similarly, when $\frac{\pi}{2}\leq\varepsilon< \pi$ holds, we have $
\dot{V}\leq -2\alpha_2 V $.  Hence the exponential synchronization
rate is no less than $\alpha_2$, which completes the proof.
\end{proof}

\begin{Remark 4}
When $0\leq\varepsilon<\frac{\pi}{2}$ holds and there is no
pacemaker, i.e., $G=0$, using the average phase
$\bar\varphi=\sum_{i=1}^N\frac{\varphi_i}{N}$ as reference, we can
define the relative phase as $\xi_i=\varphi_i-\bar\varphi$. Since
$\xi^T{\bf 1}=0$ with ${\bf 1}=[1,1,\ldots,1]^T$, the constraint
$\xi^T{\bf 1}=0$ is added to the optimization
$\min\limits_\xi\left\{ \xi^T \left(\sigma_1G+\sigma_2BWB^T\right)
\xi /(\xi^T\xi)\right\}$ in (\ref{eq:alpha_1}). Given that $G=0$ and
$BWB^T$ is the Laplacian matrix of interaction graph and hence has
eigenvector ${\bf 1}$ with associated eigenvalue $0$ \cite{horn:85},
$\lambda_{\min}$ in (\ref{eq:alpha_1}) reduces to the second
smallest eigenvalue, which is the same as the convergence rate in
section IV of \cite{Chung:10} obtained using contraction analysis.
\end{Remark 4}

Eqn. (\ref{eq:alpha_2}) shows that when
$\max\limits_{i}|\xi_i|=\varepsilon\geq\frac{\pi}{2}$, a stronger
pacemaker, i.e., a larger $g_{\min}$ leads to a larger $\alpha_2$,
but the relation between $\alpha_1$ and $g_i$
 when $\max\limits_{i}|\xi_i|=\varepsilon<\frac{\pi}{2}$ is
not clear. (In this case, $g_{\min}$ may be zero.) We can prove that
in this case $\alpha_1$ also increases with $g_i$ for any
$i=1,2,\ldots,N$:
\begin{Theorem 3}\label{theo:Theorem 3}
Both $\alpha_1$ in (\ref{eq:alpha_1}) and $\alpha_2$ in
(\ref{eq:alpha_2}) increase with an increase in pacemaker strength.
\end{Theorem 3}
\begin{proof}
As analyzed in the paragraph above Theorem \ref{theo:Theorem 3}, we
only need to prove Theorem \ref{theo:Theorem 3}
 when $\varepsilon<\frac{\pi}{2}$ holds, i.e., $\alpha_1$ is an
increasing function of $g_i$. Recall from (\ref{eq:explicit
formulation of Lambda}) that $\sigma_1G+\sigma_2BWB^T$ is an
irreducible matrix with non-positive off-diagonal elements, so there
exists a positive $\mu$ such that $\mu I-(\sigma_1G+\sigma_2BWB^T)$
is an irreducible non-negative matrix. Therefore,
$\lambda_{\max}\left(\mu I-(\sigma_1G+\sigma_2BWB^T)\right)$ is the
Perron-Frobenius eigenvalue of $\mu I-(\sigma_1G+\sigma_2BWB^T)$ and
is positive \cite{horn:85}. Given that for any $1\leq i\leq N$,
$\mu-\lambda_i(\sigma_1G+\sigma_2BWB^T)$ is an eigenvalue of $\mu I
-(\sigma_1G+\sigma_2BWB^T)$ where $\lambda_i$ denotes the $i$th
eigenvalue, we have $
\mu-\lambda_{\min}(\sigma_1G+\sigma_2BWB^T)=\lambda_{\max}\left(\mu
I -(\sigma_1G+\sigma_2BWB^T)\right) $,  i.e., $
\alpha_1=\lambda_{\min}(\sigma_1G+\sigma_2BWB^T)=\mu-\lambda_{\max}\left(\mu
I -(\sigma_1G+\sigma_2BWB^T)\right)$.

Since the Perron-Frobenius eigenvalue of $\mu
I-(\sigma_1G+\sigma_2BWB^T)$ is an increasing function of  its
diagonal
 elements \cite{horn:85}, which are decreasing functions of all $g_i$, it follows  that $\lambda_{\max}\left(\mu I
-(\sigma_1G+\sigma_2BWB^T)\right)$ is a decreasing function of
$g_i$, meaning that $\alpha_1$ is an increasing function of all
$g_i$.
\end{proof}

\begin{Remark 2}
When all $\xi_i$  are in $(-\frac{\pi}{2},\,\frac{\pi}{2})$, since
$S_2$ in (\ref{eq:S1 and S2}) is positive definite, which leads to
$-\xi^TBWS_2B^T\xi< 0$, the local coupling will increase $\alpha_1$
in (\ref{eq:alpha_1}). But when $\max\limits_{i}|\xi_i|$ is larger
than $\frac{\pi}{2}$,  $S_2$  can be indefinite, hence
$-\xi^TBWS_2B^T\xi$ can be positive, negative or zero, thus the
local coupling may increase, decrease or have no influence on the
synchronization rate. This conclusion is confirmed by
 simulations in Sec. \ref{sec:numerical simulation}.
\end{Remark 2}
\section{The non-identical natural
frequency case}\label{se:synch rate under non-identical natural
frequencies} When natural frequencies  are non-identical, Kuramoto
oscillators cannot be fully synchronized \cite{acebron:05,
Strogatz:00}. Next, we will prove that synchronization can be
achieved in the sense that the synchronization error (defined as the
maximal relative phase) can be made arbitrarily small. This is done
in two steps: first we show that  under some conditions, the
oscillators can be phase-locked, then we prove that the relative
phases can be trapped in $[-\delta,\,\delta]$ for an arbitrary
$\delta>0$ if the pacemaker is
 strong enough. The role played by the phase trapping approach
is twofold: on the one hand, it makes the conditions required in
phase locking achievable, and on the other hand, in combination with
the phase locking, it can reduce the phase synchronization error to
an arbitrary level.

\subsection{Conditions for phase locking}\label{sec:condition for frequency
synchronization} When the natural frequencies  are non-identical,
the dynamics of the oscillator network are given in (\ref{eq:matrix
form}). As in previous studies, we assume that the natural
frequencies are constant with respect to time. The results are
summarized below:
\begin{Theorem 4}\label{theo:Theorem 4}
Denote $\varepsilon\triangleq\max\limits_{i}|\xi_i|$, then the
network in (\ref{eq:matrix form}) can achieve phase locking if
\begin{enumerate}
\item  $0\leq
\varepsilon<\frac{\pi}{4}$ holds, at least one $g_i$ is positive,
and the coupling $a_{i,j}$ is connected;
\item $\frac{\pi}{4}\leq \varepsilon< \frac{\pi}{2}$ and
$
g_{\min}>\bigg\{\frac{-\cos(2\varepsilon)\lambda_{\max}(BWB^T)}{\cos\varepsilon},\:
\max\limits_i\big\{\sum\limits_{j=1,j\neq
i}^N-\frac{a_{i,j}\cos(2\varepsilon)}{\cos\varepsilon}\big\}\bigg\}
$ hold.
\end{enumerate}
\end{Theorem 4}
\begin{proof}
To prove phase locking, i.e., all oscillators oscillate at the same
frequency, we need to prove that the oscillating frequencies
$\dot\varphi_i$ are identical. From (\ref{eq:xi}), we have
$\dot\varphi_i=w_0+\dot\xi_i$, so if $\zeta\triangleq\dot{\xi}$
converges to zero, then phase locking is achieved.

Differentiating (\ref{eq:matrix form}) yields
\begin{equation}\label{eq:dot Xi}
\dot{\zeta}=-GS_3\zeta-BWS_4B^T\zeta
\end{equation}
where
\begin{eqnarray}\label{eq:S_4 in theorem 5}
\begin{aligned}
S_3&=\textrm{diag}\left(\cos\xi_1,\,\cos\xi_2,\,\ldots,\,\cos\xi_N\right),\\
S_4&=\textrm{diag}\left(\cos(B^T\xi)_1,\,\cos(B^T\xi)_2,\,\ldots,\,\cos(B^T\xi)_M\right)
\end{aligned}
\end{eqnarray}

Following the line of reasoning of the proof of Theorem
\ref{theo:Theorem 1}, we can prove that  $\zeta$ is positively
invariant under conditions in Theorem \ref{theo:Theorem 4}. Next we
proceed to prove the convergence of $\zeta$.

Define a Lyapunov function as $ V=\frac{1}{2}\zeta^T\zeta$.
Differentiating $V$ along the trajectory of (\ref{eq:dot Xi}) yields
\begin{equation}\label{eq:vdot in theorem 5}
\dot{V}=\zeta^T\dot{\zeta}=-\zeta^TGS_3\zeta-\zeta^TBWS_4B^T\zeta
\end{equation}

Following the line of reasoning of  Theorem \ref{theo:Theorem 1},
when $\zeta\neq0$, we can obtain  $\dot{V}<0$ under the conditions
in Theorem \ref{theo:Theorem 4}. So $V$, and hence  $\zeta$ will
converge to 0. Thus oscillating frequencies become identical and
 phase locking is achieved.
\end{proof}

\begin{Remark 5}
In the absence of a pacemaker, the authors in \cite{chopra:09}
proved that if the phase difference between any two oscillators,
i.e., $\varphi_i-\varphi_j, \,\forall i,j$, is within
$[-\frac{\pi}{2},\,\frac{\pi}{2}]$, then phase locking can be
achieved. Given $\varphi_i-\varphi_j=\xi_i-\xi_j, \,\forall i,j$,
the
 condition in \cite{chopra:09} only applies to $-\frac{\pi}{4}\leq \xi_i\leq
\frac{\pi}{4}$ in our formulation framework.  
\end{Remark 5}

\subsection{A bound on the exponential rate of phase locking}\label{se:a bouned on the exponential
synchronization rate of frequency synchronization}

\begin{Theorem 5}\label{theo:Theorem 5}
For the network in (\ref{eq:matrix form}), denote
$\varepsilon=\max\limits_i|\xi_i|$. If the conditions in Theorem
\ref{theo:Theorem 4} are satisfied, then
\begin{enumerate}
\item when
$0\leq \varepsilon< \frac{\pi}{4}$ holds,  the exponential
phase-locking rate is no worse than
\begin{equation}\label{eq:alpha_3}
\begin{aligned}
\alpha_3&
\lambda_{\min}\left(\sigma_3G+\sigma_4BWB^T\right)
\end{aligned}
\end{equation}
with $\sigma_3\triangleq\cos\varepsilon$ and $
\sigma_4\triangleq\cos2\varepsilon$;
\item when
$\frac{\pi}{4}\leq\varepsilon< \frac{\pi}{2}$ holds, the exponential
phase-locking rate is no worse than \vspace{-0.2cm}
\begin{eqnarray}\label{eq:alpha_4}
\alpha_4=g_{\min}{\cos}(\varepsilon)+{\cos}(2
\varepsilon)\lambda_{\max}(BWB^T)
\end{eqnarray}
\end{enumerate}
\end{Theorem 5}
\begin{proof}
 Theorem
\ref{theo:Theorem 5} can be derived following the line of reasoning
of  Theorem \ref{theo:Theorem 2} and thus is omitted.
\end{proof}
\begin{Remark 7}
Following Theorem \ref{theo:Theorem 3}, we can prove that a stronger
pacemaker always increases $\alpha_3$ (and $\alpha_4$). But a
stronger local coupling can have different impacts: when $0\leq
\varepsilon<\frac{\pi}{4}$, $S_4$ in (\ref{eq:S_4 in theorem 5}) is
positive definite, $-\zeta^TBWS_4B^T\zeta$ is negative, so the local
coupling will increase $\alpha_3$. However, when $\frac{\pi}{4}\leq
\varepsilon< \frac{\pi}{2}$, since $S_4$ in (\ref{eq:S_4 in theorem
5}) can be indefinite, $-\zeta^TBWS_4B^T\zeta$ can be positive or
negative. Thus the local coupling may increase or decrease the rate
of phase locking. The conclusion will be confirmed by simulations in
Sec. \ref{sec:numerical simulation}.
\end{Remark 7}
\subsection{Method for trapping relative phases}\label{se:phase
trapping} In this section, we will give a method such that the
relative phases are trapped in any interval $[-\delta,\,\delta]$
with an arbitrary $0<\delta< \pi$. 
\begin{Theorem 6}\label{theo:Theorem 6}
For (\ref{eq:matrix form}) with frequency differences $\Omega$,
denote $\varepsilon=\max\limits_i|\xi_i|$ and
$\|\Omega\|=\sqrt{\Omega^T\Omega}$, then the relative phases can be
trapped in a compact set $[-\delta,\,\delta]$ for an arbitrary
$0<\delta< \pi$
\begin{enumerate}
\item if $0\leq\varepsilon<
\frac{\pi}{2}$  and the following condition is satisfied:
\begin{equation}\label{eq:condtion 1 in theorem 7}
g_{\min}> \|\Omega\|/(\delta{\rm sinc}(\varepsilon))
\end{equation}
\item if
$\frac{\pi}{2}\leq\varepsilon< \pi$  and the following condition is
satisfied:
\begin{equation}\label{eq:condition 3 in theorem 7}
\hspace{-0.25cm}g_{\min}> \|\Omega\|/(\delta{\rm
sinc}(\varepsilon))-\frac{{\rm
sinc}(2\varepsilon_0)\lambda_{\max}(BWB^T)}{{\rm sinc}(\varepsilon)}
\end{equation}
where $\varepsilon_0$ is defined in (\ref{eq:tilde}).
\end{enumerate}
\end{Theorem 6}
\begin{proof}
Differentiating Lyapunov function $ V=\frac{1}{2}\xi^T\xi $ along
the trajectory of (\ref{eq:matrix form}) yields
\begin{equation}\label{eq:V in theorem 7}
\begin{aligned}
\dot{V}=\xi^T\dot{\xi}&=\xi^T\Omega-\xi^TG\sin\xi-\xi^TBW\sin(B^T\xi)\\
&=\xi^T\Omega-\xi^TGS_1\xi-\xi^TBWS_2B^T\xi
\end{aligned}
\end{equation}
with $S_1$ and $S_2$ defined in (\ref{eq:S1 and S2}).
\begin{enumerate}
\item When  $0\leq\varepsilon<
\frac{\pi}{2}$ holds, we have $S_1\geq {\rm sinc}(\varepsilon)I>0$
and  $S_2\geq 0$ from previous analysis.
 Using (\ref{eq:V in theorem 7}),
 (\ref{eq:S1 and S2}), and the fact $\lambda_{\min}(G)=g_{\min}$,
we have
\begin{equation}
\begin{aligned}
\dot{V}&\leq \|\xi\|\|\Omega\|-g_{\min} {\rm
sinc}(\varepsilon)\|\xi\|^2
\end{aligned}
\end{equation}
If $\xi_i$ is outside $[-\delta, \, \delta]$ for some $i$, we have
$\|\xi\|=\sqrt{\sum_{i=1}^N\xi_i^2}>\delta$, which in combination
with (\ref{eq:condtion 1 in theorem 7}) leads to $\dot{V}<0$.
Therefore all $\xi_i$ will converge to $[-\delta, \, \delta]$.
\item When
$\frac{\pi}{2}\leq \varepsilon<\pi$ holds, from the analysis in
Theorem \ref{theo:Theorem 1}, we have
 $
S_1\geq {\rm sinc}(\varepsilon)I>0$ and $S_2\geq {\rm
sinc}(2\varepsilon_0)I$. Then using  (\ref{eq:V in theorem 7}) and
the fact $\lambda_{\min}(G)=g_{\min}$, we have
\begin{equation}
\begin{aligned}
\dot{V}&\leq \|\xi\|\|\Omega\|-g_{\min}{\rm
sinc}(\varepsilon)\|\xi\|^2\\
&\quad -{\rm sinc}(2\varepsilon_0)\lambda_{\max}(BWB^T)\|\xi\|^2
\end{aligned}
\end{equation}
If $\xi_i$ is outside  $[-\delta, \, \delta]$ for some $i$, we have
$\|\xi\|>\delta$, which in combination with (\ref{eq:condition 3 in
theorem 7}) leads to $\dot{V}<0$. Thus all $\xi_i$ will converge to
the interval $[-\delta, \, \delta]$.
\end{enumerate}
\vspace{-0.5cm}
\end{proof}
\begin{Remark 10}
Theorem \ref{theo:Theorem 6} used the important fact that if
$\|\xi\|=\sqrt{\sum_{i=1}^N \xi_i^2}$ is restricted to the interval
$[0,\,\delta]$, then all $\xi_i$ are restricted to the interval
$[-\delta,\,\delta]$.
\end{Remark 10}
\begin{Remark 8}
When $\|\xi\|<\pi$, \cite{chopra:09} gives a condition under which
$\xi_i$ can be trapped in an arbitrary compact set. Since for a
large number of oscillators $N$, $\|\xi\|=\sqrt{\sum_{i=1}^N
\xi_i^2}\leq \pi$ is difficult to satisfy, our result is more
general.
\end{Remark 8}


\section{Simulation results}\label{sec:numerical simulation}
We consider a network composed of $N=9$  oscillators. The coupling
strengths $a_{i,j}$ are randomly chosen from the interval $[0,\,
0.1]$. They were found to form a connected interaction graph.  As in
previous studies, we use the modulus of the order parameter
$r=\left|\frac{1}{N}\sum_{i=0}^{N}e^{j\varphi_i}\right| $ to measure
the degree of synchrony \cite{Strogatz:00}. The value of $r$
($r\in[0,\:1]$) will approach $1$ as the network is perfectly
synchronized, and $0$ if the phases are randomly distributed
\cite{Strogatz:00}.
According to \cite{Strogatz:00}, we have $r\, \approx 1$ when the
oscillators are synchronized. So  we define synchronization to be
achieved when $r$ exceeds $0.99$.

When the natural frequencies are identical, we set the phase of the
pacemaker $\varphi_0$ to $\varphi_0=w_0t$ with $w_0=1$ and simulated
the network using initial phases $\varphi_i=\varphi_0+\xi_i$ with
$\xi_i\in(-\frac{\pi}{2},\,\frac{\pi}{2})$ and initial phases
$\varphi_i=\varphi_0+\xi_i$ with
 $\xi_i\in(-\pi,\,\pi)$, respectively.  In the former case, we
connected the first oscillator to the pacemaker and set $g_1=g,\,
g_2=g_3=\ldots=g_9=0$. In the latter case, we connected all
oscillators to the pacemaker and set $g_1=g_2=\ldots=g_9=g$. In both
cases, we set $g=1$. To show the influences of the pacemaker on the
synchronization rate, we fixed $a_{i,j}$ and simulated the network
under different  pacemaker strengths $m\times g$, where
$m=1,2,\ldots,10$. To show the influences of local coupling on the
synchronization rate, we fixed the strength of the pacemaker to $3g$
and simulated the network under different local coupling strengths
$m\times a_{i,j}$ for all $a_{i,j}$, where $m=1,2,\ldots,10$. All
the synchronization times are averaged over 100 runs with initial
$\xi_i$ in each run randomly chosen from a uniform distribution on
$(-\frac{\pi}{2},\,\frac{\pi}{2})$ (in the former case) or on
$(-\pi,\,\pi)$ (in the latter case). The results are given in Fig.
1. It is clear that a stronger pacemaker always increases the
synchronization rate, whereas the local coupling increases the
synchronization rate when all $\xi_i$ are within
$(-\frac{\pi}{2},\,\frac{\pi}{2})$, and it may increase or decrease
the synchronization rate when the maximal/minimal $\xi_i$ is outside
$(-\frac{\pi}{2},\,\frac{\pi}{2})$.
 \begin{figure}[!hbp]
\vspace{-0.5cm}
       \includegraphics[width=\columnwidth]{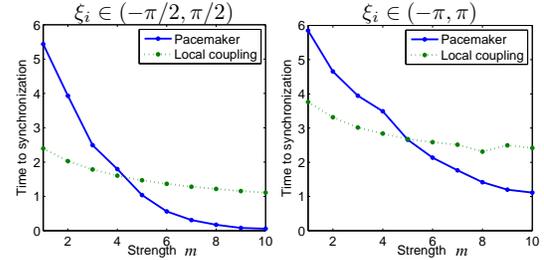}
       \vspace{-0.6cm}
    \caption{Times to synchronization under different strengths of
pacemaker/local coupling (with all oscillators having identical
natural frequencies).}
\vspace{-0.4cm}
\end{figure}
\begin{figure}[!hbp]
\vspace{-0.5cm}
        \includegraphics[width=\columnwidth]{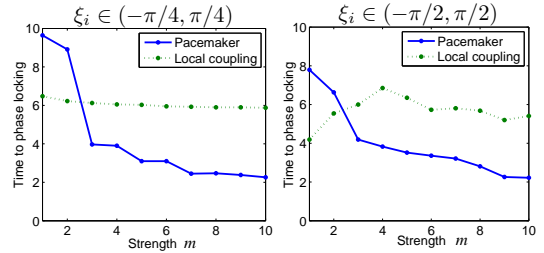}
               \vspace{-0.6cm}
    \caption{Times to phase locking under different strengths of
pacemaker/local coupling (with oscillators having non-identical
natural frequencies).}
\vspace{-0.1cm}
\end{figure}


When the natural frequencies  are non-identical, we simulated the
network using initial  phases $\varphi_i=\varphi_0+\xi_i$ with
$\xi_i\in(-\frac{\pi}{4},\,\frac{\pi}{4})$ and initial phases
$\varphi_i=\varphi_0+\xi_i$ with
$\xi_i\in(-\frac{\pi}{2},\,\frac{\pi}{2})$, respectively. In the
former case, we connected the first oscillator to the pacemaker and
set $g_1=g$. In the latter case, we connected all the oscillators to
the pacemaker and set $g_1=g_2=\ldots=g_9=g$. The natural
frequencies were randomly chosen from  $(0,\,1)$. Tuning the
strengths in the same way as in the identical natural frequency
case, we simulated the network  under different strengths of the
pacemaker and local coupling. All of the times to phase locking
 are averaged over 100 runs with initial $\xi_i$
 randomly chosen from a uniform distribution on
$(-\frac{\pi}{4},\,\frac{\pi}{4})$ (in the former case) or on
$(-\frac{\pi}{2},\,\frac{\pi}{2})$ (in the latter case). The results
are given in Fig. 2. It is clear that a stronger pacemaker always
increases the rate to phase locking, whereas the local coupling
increases the rate to phase locking when all $\xi_i$ are within
$(-\frac{\pi}{4},\,\frac{\pi}{4})$, and it may increase or decrease
the rate to phase locking when the maximal/minimal $\xi_i$ is
outside $(-\frac{\pi}{4},\, \frac{\pi}{4})$.

 To confirm the prediction that $\xi_i$  can be made smaller by
 making the pacemaker strength stronger, we set $g_1=\ldots=g_9=g$ and simulated the network under initial phases
  $\varphi_i=\varphi_0+\xi_i$ with
$\xi_i\in(-\pi,\,\pi)$. Using the same $\xi_i$, the maximal final
relative phase when the strength of the pacemaker $g$ is made $m$
$(m=1,2,\dots,10)$ times greater is recorded and  given in Fig.
\ref{fg:phase
 trapping.eps}. It can be seen that the maximal final relative phase
 (i.e., synchronization error) decreases with the strength of the pacemaker, confirming the prediction in Theorem \ref{theo:Theorem 6}.
 \begin{figure}[!hbp]
 \vspace{-0.2cm}
\begin{center}
        \includegraphics[width=0.66\columnwidth]{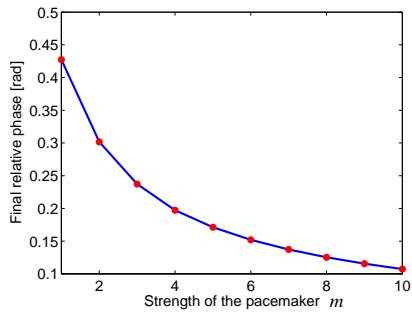}
    \caption{The maximal final relative phase (phase synchronization error) under different strengths of the pacemaker when  oscillators have non-identical natural frequencies (which are randomly chosen from the interval $(0,\,1)$).}
    \label{fg:phase trapping.eps}
\end{center}
\vspace{-0.5cm}
\end{figure}

\section{Conclusions}\label{se:conclusions}
The  exponential synchronization rate of  Kuramoto oscillators is
analyzed in the presence of a pacemaker. In the identical natural
frequency case, we prove that synchronization to the pacemaker can
be ensured  even when the initial phases
 are not constrained in an open half-circle, which improves the existing results in the literature. Then we derive a lower bound on the exponential synchronization rate, which is
 proven
 an increasing function of the  pacemaker strength, but may be an
 increasing or decreasing function of the local
 coupling strength. In the non-identical natural frequency case,
  a similar conclusion is
 obtained on phase locking. In this case, we also prove that relative phases (synchronization error) can be made
 arbitrarily small by making the pacemaker strength strong enough. The results are independent of oscillator numbers in the
 network and are confirmed by numerical simulations.

\bibliographystyle{unsrt}
\bibliography{abbr_bibli}

\end{document}